\documentclass[iop,author-year]{emulateapj}
\usepackage{url}
\usepackage{graphicx}
\usepackage{color}
\usepackage{amsmath}
\usepackage{setspace}
\usepackage{threeparttable}
\usepackage{epstopdf}
\usepackage{graphics}
\usepackage{subfigure}
\usepackage{url}             
\usepackage{rotating}
\usepackage{lipsum}
\usepackage{float}
\usepackage{wrapfig}
\usepackage{longtable}
\usepackage{natbib}
\usepackage{hyperref}
\renewcommand{\thefootnote}{\ifcase\value{footnote}\or * \or \# \or $\dagger$ \or ** \or \# \or * \or \# \or $\dagger$ \or ($\infty$)\fi}

\shorttitle{Astronomical Catalogs for Locating Gravitational-wave Events}
\shortauthors{Kunyang Li et al.}

\begin{document}

\title{Astronomical Catalogs for Locating Gravitational-wave Events}

\author{Kunyang Li\altaffilmark{1}, \and Roy D.Williams\altaffilmark{2}}
\affil{Department of Physics and Space Sciences, 150 W. University Blvd., Florida Institute of Technology, Melbourne, FL 32901, USA}
\affil{LIGO, California Institute of Technology, Pasadena, CA 91125, USA}

\email{lik2013@my.fit.edu}

\begin{abstract}
Gravitational wave transients are caused by some of the most energetic events in the Universe, and a precise location would allow deep examination of the counterpart by electromagnetic waves (telescopes collecting light), the combination of GW and EM resulting in very much improved science return (multi-messenger Astronomy). Since the GW detectors do not provide good localization on the sky, the faint counterpart will be very difficult to find. One strategy to help the search is to look first where mass is concentrated, and thus the prior probability of GW events is highest. In the first part of this paper, we present methods used to estimate stellar masses and metallicities of galaxies and galaxy clusters in  different catalogues. In the second part of the paper, we test our estimation accuracy by comparing our results with stellar masses given in Stripe 82 Massive Galaxy Catalogue (S82-MGC). The relation between stellar mass we found and that from S82-MGC is provided for GWGC, 2MASS-GLADE, and WISExSCOS catalog in the last part of the paper. Our results are used in an interactive web-based tool (Skymap Viewer) for astronomers to decide where to look first in EM follow-up observations of GW events in the future.


\end{abstract}


\section{Introduction}

The first gravitational wave detection in September 2015 was a magnificent achievement, proving the existence of black holes of ~30 solar masses, proving that such objects can inspiral and coalesce, and proving that Einstein's original conception of gravity is correct even in these strong, dynamic gravitational fields.

But great further understanding would be possible if conventional electromagnetic astronomy (optical, X-ray, radio, etc, collectively ``EM'') could be brought to bear on the gravitational-wave (GW) source. As with the flowering of insight following optical identification of gamma-ray bursts in the 1990's, we hope that a program of rapid follow-up will be able to identify the EM counterpart of a future GW source, and for this multi-messenger astronomy to lead to greater insight. 

The GW detection by the two LIGO instruments has not provided a tight localization in the sky, but rather a broad probability distribution (the ``skymap'') covering hundreds of square degrees on the sky. As more detectors come online in the next few years (Italy, Japan, India), we expect the localization to be much tighter, and the search area much smaller.

Presumably the GW source is located in a galaxy, where almost all the matter resides, and so the best place to look for the EM counterpart is among those galaxies that are in the skymap, with a priority given to more massive galaxies. Presumably the more matter that is there, the better chance of black-hole binaries being present. Earlier runs of LIGO from 2010 hoped to detect sources at a distance up to 100 Mpc, and a galaxy catalog was built (GWGC) listing about 60,000 galaxies out to that distance.

However, the LIGO detection was at a distance of 400 Mpc, four times further. Galaxy catalogs are incomplete out to this distance, and of course the numbers are much larger -- if distance is four times further, then galaxy numbers multiply by 64, proportional to volume. However in the future, we expect there to be more GW detectors operating, with a much better localization of the source, so fewer galaxies in the observing list.

\subsection{Skymap Viewer}
\begin{figure}[htp]
\centering
\includegraphics[width=9cm]{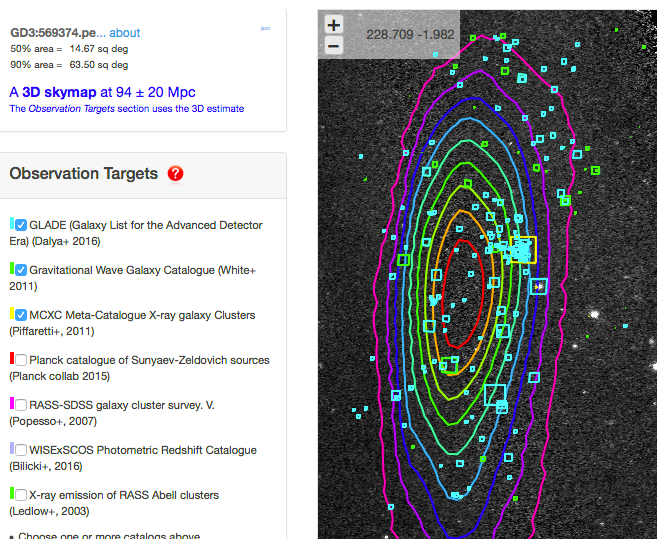}
\caption{Screenshot of Skymap Viewer\citep{skymap}, with contours of the skymap, overlaid by selected catalogs of astronomical sources (here ROSAT All-Sky Survey), with the area of each square proportional to the mass of the corresponding source.}
\end{figure}

 Skymap Viewer \citep{skymap} is an interactive, web-based tool to display a skymap along with a host of relevant information for follow-up observers. The skymap is shown as a contour plot, each color-coded line enclosing a given percentage of the total probability. The AladinLite platform \citep{aladin} , on which Skymap Viewer is built, enables drill-down from whole-sky to arc-second resolution, including image surveys from radio to gamma-ray wavelengths. Also shown are the sun, moon, and milky way relative to the skymap, at any user-chosen time and location on the Earth.
 
This paper defines a new capability: to visualize arbitrary astronomical catalogs in terms of {\it observation priority},  the posterior, which combines new data from the gravitational wave detection (the skymap), with known astrophysical objects, the prior. Skymap Viewer thus shows the product of these two terms: 
\begin{quote} 
\centering 
   Observation priority = \\GW skymap $\times$ astrophysical prior
\end{quote}
where the astrophysical prior is based on stellar mass or blue luminosity, with future extensions to include, for example, metallicity. However, rather than building a single ``expert'' view of the astrophysical prior, we will provide the ability for a scientist to utilize the catalogs they judge most important: catalogs of galaxies, or of galaxy clusters, or of X-ray sources, for example.

\subsection{Astrophysical Prior}
The focus of this paper is thus to compute astrophysical priors from attributes given in object catalogs. This in turn depends on theoretical ideas about environments that encourage binary neutron star systems (BNS), or binary black hole systems (BBH). Clearly mass is crucial: the more stars there are, the higher the chance of exotic systems that create GW. However, it is thought that BNS is a result of recent/current star-forming activity \citep{BNS} , and therefore that the mass of \textit{blue} stars (starburst and young stars) might be a better proxy than the overall mass of stars. Note however, that the binary may have existed for bllions of years before merging, so the starburst that created the neutron stars may be long gone. The BBH mergers detection range of aLIGO (advanced LIGO) reaches to 1000 Mpc, but the BNS mergers detection range is more like 100 Mpc, due to the smaller masses of BNS mergers. So both stellar mass and star-formation rate also contribute to the priors of galaxies and galaxy clusters closer than $ \sim 100 $ Mpc, but for more distant ones, the prior is dominated by  stellar masses.

\section{Galaxy Mass Estimation}

We summarize three catalogs of galaxies and the methods we have used to estimate stellar mass.

\subsection{Estimating mass with blue magnitude for GWGC catalog}
For the GWGC catalog, each source has a distance estimate as well as a blue magnitude (Bmag), so we could try to compute the galaxy stellar mass from it. However, blue light can under-emphasize mass as blue light is easily obscured by circumstellar material. A very simple mass estimator is to assume that the ratio of mass to blue luminosity is the same as the Sun, leading to:
\begin{equation*}
    M = 0.00855 d^2 10^{-0.4*Bmag}
\end{equation*}
where $d$ is Mpc and $M$ is in units of $10^{14} M_{\sun}$. We will test this simple rule in section 3.1.

\subsection{Bayesian SED Fitting for the GLADE catalog}
The GLADE catalog has been combined and matched from four existing galaxy catalogs: GWGC, 2MPZ, 2MASS XSC and HyperLEDA. GLADE associated B-band magnitudes and photometric redshifts for 548,876 2MASS \citep{2MASS} galaxies which lacked these properties with a machine-learning algorithm whose training data was a subsample of the 2MPZ catalog \citep{2mpz}. Thus each galaxy has infreared (JHK) magnitudes, as well as Blue magnitude. Using these, we can estimate galaxy mass by Bayesian SED fitting, using the evolutionary population synthesis model by Bruzual and Charlot \citep{bc2003}. The BayeSED code\citep{han} estimates stellar mass and metallicity for each galaxy in the GLADE catalog. For the BC2003 model used in BayeSED, the IMF of Chabrier \citep{imf} is adopted, and the star-formation histories of galaxies are assumed to be exponentially declining at rate $\tau$. A uniform screen geometry is combined with the dust extinction estimate \citep{screen}.The BC2003 model library used in BayeSED covers a parameter grid with $log( \tau /yr)$  ranging from 6.5 to 11 in steps of 0.10, log(age/yr) ranging from 7.0 to 10.1 in steps of 0.05, Av ranging from 0 to 4 in steps of 0.2, and 4 different metallicities: 0.004, 0.008, 0.02, or 0.05. In total, BayeSED offers 243,434 model SEDs in the library, which we used to compare with observed galaxy SEDs. This computation has taken approximately 3 days for all 1,498,920 GLADE galaxies.

\subsection{W Band Photometry for the WISE$\times$SCOS Catalogue }

WISE$\times$SCOS (WISE x SuperCOSMOS Photometric Redshift Catalog)\citep{wisexscos} is constructed by cross-matching between two largest all-sky photometric catalogs, mid-infrared WISE\citep{w} and SuperCOSMOS all sky samples\citep{s} and applying the artificial neural network approach\citep{2004}, trained on the GAMA-II redshift survey \citep{d}\citep{2015}. The derived photometric redshifts have errors nearly independent of distance, with an overall accuracy of $\sigma $z/(1+z) = 0.033, a typical precision of 15$\% $. The WISE$\times$SuperCOSMOS catalog provides (after applying the cleaning mask) about 18.5 million galaxies with a median redshift of $z$=0.2, and a sky coverage of 3$\pi $ steradians. Only those sources that have photometry data in at least four bands: W1 and W2 (3.4 and 4.6 $\mu$m) in WISE, B and R in SCOS are selected into WISE$\times$SCOS from these two catalogs. 

The W1 band of WISE is dominated by the light from old stars and can be used as an effective measure of galaxy stellar mass \citep{w1}\citep{wen}. According to Cluver et al. \citep{w1}, the empirical relation between optically determined stellar mass (estimated from SPS models) and W1, W2 WISE measurements is:
\begin{equation*}
    \log_{10}(M_{stellar}/L_{W1})=-1.96(W_1-W_2)-0.03
\end{equation*}
\begin{equation*}
    L_{W1} (L_\sun) = 10^{-0.4(M - M_\sun)},
\end{equation*}
where M is absolute magnitude in W1, $ M_\sun=3.24$, and $ W_{1}-W_{2}$ is observed color. We have associated optical stellar mass to each galaxy in WISE$\times$SCOS by using W1, W2, redshift ($z$), and the equation in \citep{w1}.

\section{Comparing Galaxy Mass Estimations}
Next we take the mass estimates from the previous section and compare them in a small area of sky with very accurate mass estimates. This small area is ``Stripe 82'',  a region of 300 $deg^2$ on the celestial equator that the Sloan Digital Sky Survey (SDSS) has imaged over 10 times, giving co-added optical data 2 times deeper. The ``S82 Massive Galaxy Catalog'' \citep{s82} has been extracted from Stripe 82, and provides the mass estimates that we use to test and calibrate the algorithms described above.

\subsection{Comparison between the GWGC catalog and S82-MGC}
To test how well B band photometry data tell us about the stellar mass of a galaxy, we have built a cross-match of GWGC with S82-MGC \citep{s82}. Using this cross-match, we have compared the stellar mass precisely estimated by SED fitting in S82-MGC with that derived from the blue luminosity in GWGC, as shown in Fig. 2.

\begin{figure}[htp]
\centering
\includegraphics[width=9cm]{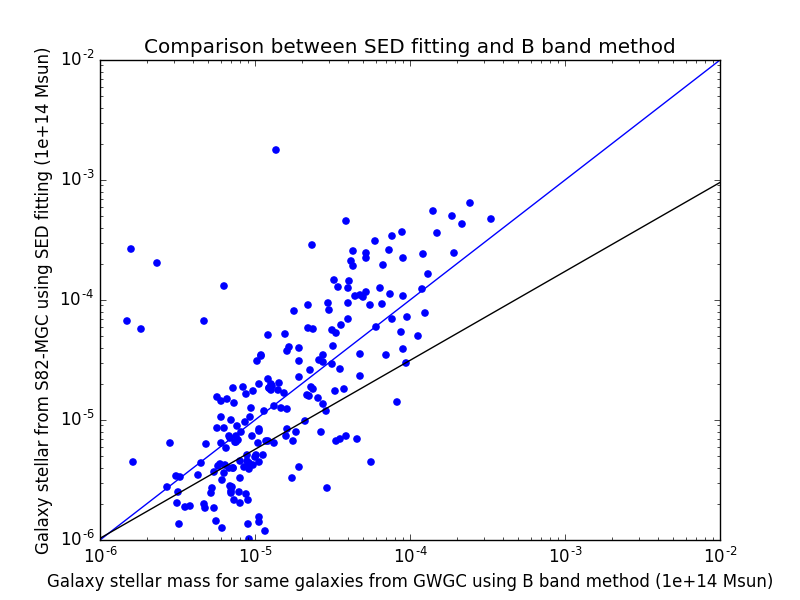}
\caption{log(Blue Mass) - log(Stellar Mass) Plot for Cross-matched GWGC galaxies}
\end{figure}

The best fit line in Fig. 2 shows the relation between GWGC mass and S82-MGC mass for the same group of galaxies,

\begin{equation}
    \log_{10}(M_{S82})=0.7414\log_{10}(M_{GWGC})-1.538,
\end{equation}

where $M_{GWGC}$ is mass estimated using B band method and GWGC data. More data is needed to confirm this relation.

\subsection{Comparison between the GLADE Catalog and S82-MGC}

We have taken a subset of GLADE that has both B magnitude and the JHK magnitudes from 2MASS. With these, we have estimated mass using BayeSED code and compared with those provided in S82-MGC using the same method but different photometry data. 
To precisely estimate the stellar mass of a galaxy, stellar population synthesis (SPS) modelling and SED fitting are needed.
S82-MGC provides relatively precise stellar mass estimated by Bayesian SED fitting between Y JHK photometry from the UKIDSS Large Area Survey (LAS) \citep{ukidss} and FSPS models (FSPS: Flexible Stellar Population Synthesis \citep{conroy09} , 2010 \citep{conroy10}). We performed a cross matching between GLADE and S82-MGC to calibrate how well the blue luminosity describes the stellar mass of a galaxy in GLADE. 

To find the best search radius of this cross-matching, we looped through GLADE galaxies inside the sky coverage of S82-MGC using a range of search radius (rs) from 0.0005 to 0.05 degrees. Each time, for each galaxy in GLADE, the number of S82-MGC galaxies that fall into a rectangle centered at the GLADE galaxy with length rs is counted, and the total numbers of GLADE galaxies that have 0, 1, and 2 S82-MGC galaxies in the searching rectangle are stored into lists Nrs0, Nrs1, and Nrs2, which are of the same length of the search radius list. In the end, rs is plotted against Nrs0, Nrs1, and Nrs2. The best search radius is the point where Nrs1 curve peaks, and Nrs2 curve starts to raise from zero. We found that the best searching radius is about 0.0005 degree, which we used, getting 4787 matched galaxies. We plot in Fig. 3 the stellar mass from S82-MGC and the mass derived from SED fitting from GLADE-2MASS. 




\begin{figure}[htp]
\centering
\includegraphics[width=9cm]{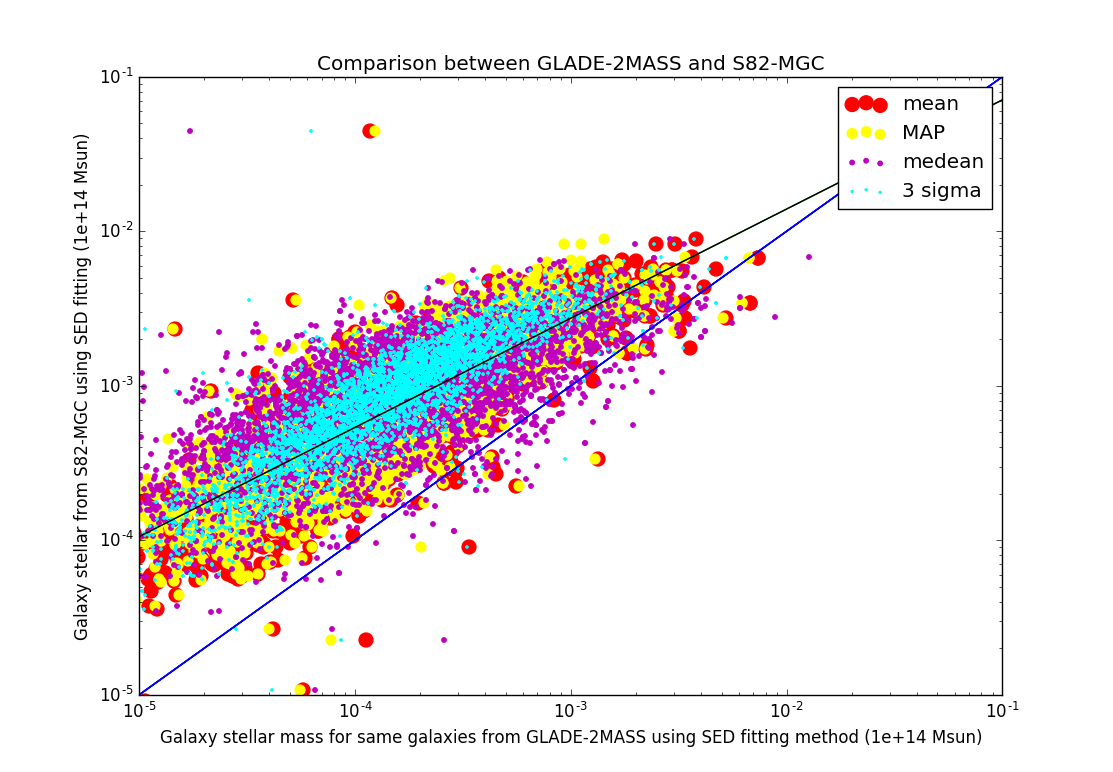}
\caption{Log-log plot of Stellar Mass Comparison Between GLADE-2MASS and S82-MGC: red dots are S82-MGC IR mass estimation, while green dots are S82-MGC optical mass estimation.}
\end{figure}
The best fit line in Fig. 3 shows the relation between GLADE-2MASS mass and S82-MGC mass for the same group of galaxies,

\begin{equation}
    \log_{10}(M_{S82})=0.71\log_{10}(M_{GLADE-2MASS})-0.44,
\end{equation}

where $M_{GLADE-2MASS}$ is mass estimated using BayeSED and GLADE-2MASS data.

The GLADE-2MASS mass and S82-MGC mass agree well, implying that the stellar mass estimated using evolutionary population synthesis and SED fitting with B JHK photometry data from GLADE-2MASS agree with those provided in S82-MGC. 

\subsection{Comparison between WISE$\times$SCOS and S82-MGC}


The cross-matching between WISE$\times$SCOS and S82-MGC is performed using the same algorithm described above. As shown in Fig. 4, the best searching radius is also ~ 0.001 degree. So we performed the cross-matching between WISE$\times$SCOS and S82-MGC using rs=0.001 deg, yielding 74,403 matched galaxies. We plotted the stellar mass from S82-MGC and the stellar mass estimation from WISExSCOS using W1 and W2 band photometry data, as shown in Fig. 4. 

\begin{figure}[htp]
\centering
\includegraphics[width=9cm]{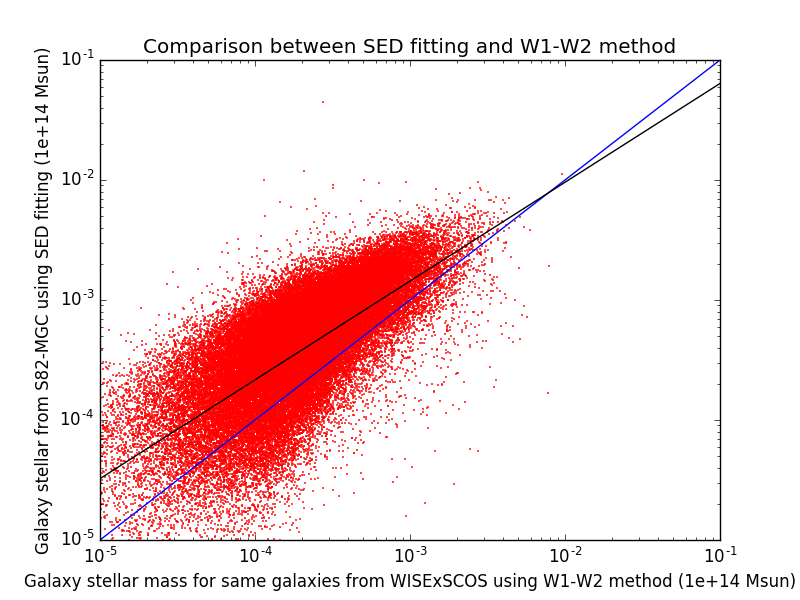}
\caption{Log-log plot of Stellar Mass Comparison Between WISExSCOS and S82-MGC}
\end{figure}

There is good agreement between the mass from WISExSCOS (section 2.2) and those provided in S82-MGC. The best fit line in Fig. 4 shows the relation between WISExSCOS mass and S82-MGC mass for the same group of galaxies,

\begin{equation}
    \log_{10}(M_{S82})=0.8233\log_{10}(M_{WISExSCOS}) -0.3723,
\end{equation}

where $M_{WISExSCOS}$ is mass estimated using $ W_{1}-W_{2}$ method and WISE$\times$SCOS data.

\section{Mass Estimation for Galaxy Cluster Catalogs}
Our objective is a listing of large accumulations of mass in the distance range of the recent LIGO discoveries (100 Mpc to 1000 Mpc), so that mass can be used as a proxy for observational priority. In this section we review the ways astronomers have used to estimate the mass of galaxy clusters.

\subsection{Cluster mass from richness}

We might directly use the richness as a mass proxy, assuming the masses of galaxy in clusters are the same. Median velocity dispersions for MaxBCG clusters binned by N200 is shown in Fig. 12 of \citep{maxbcg} , where N200 is the number of galaxies enclosed by a circle of radius R200, inside which the average density is 200 times the critical density at the corresponding redshift. The increasing velocity dispersion is related to the increasing mass of cluster in the MaxBCG catalog, according to the virial theorem, reinforcing the idea that N200 is a useful proxy for cluster mass. This method sufferes from the dubious assumption that every galaxy is of the same mass. 

Second, use the stacked velocity dispersion-richness relation derived using MaxBCG catalog data also in \citep{maxbcg} to get corresponding velocity dispersion of each cluster, and then use the virial theorem to calculate the cluster mass. 
\begin{equation*}
    M200 = \frac{5R200*\sigma(N200)^2}{G}
\end{equation*}
The best-fit power law for the relation between stacked velocity dispersion ($\sigma$) and richness (N200) is given by
\begin{equation*}
    ln{\sigma(N200)} = (5.52\pm 0.04)+(0.31\pm 0.01)ln{N200}
\end{equation*}
This method is more precise than the first method, but only if the cluster is old enough to be fully virialized. Also we still need to make an assumption that cluster radius satisfies the definition of dynamical cluster radius (R200).

Third, we can use the central halo mass-richness relation shown in Eq. 26 of \citep{sheldon} derived by applying cross-correlation cluster lensing method on SDSS data
\begin{equation*}
    M200(N200)=M200|20(N200/20)^\alpha 
\end{equation*}
with
\begin{equation*}
    M200|20=(8.8\pm 0.4\pm 1.1)\times 10^{13} h^{-1}M_\sun
\end{equation*}
\begin{equation*}
    \alpha =1.28\pm 0.04
\end{equation*}
The cluster mass can be estimated by applying the equations above on the richness N200 of each cluster. We may need to compare results from the second and the third method to see which method is more precise for our purpose. 

\subsection{Cluster mass from X-ray observations}
Galaxy clusters can be indicated by other methods than sesing the galaxies in the cluster: hot gas between the cluster members emits X-rays, and thus can be used to show the existence of a galaxy cluster.

In the case of MCXC \citep{mcxc}, the catalog lists M500 (the approximate total dynamical mass) for each cluster, which is estimated from L500, the approximate total luminosity:
\begin{equation*}
    h(z)^{-7/3}\frac{L500 10^{44} erg s^{-1}} = C(\frac{M500}{3\times 10^{14} M_\sun})^{\alpha}
\end{equation*}
\begin{equation*}
    log(C)=0.274
\end{equation*}
\begin{equation*}
    \alpha = 1.64
\end{equation*}
The adopted $C$ and $\alpha$ values are derived from REXCESS luminosity data uncorrected for the Malmquist bias, hence, the M500 provided by MCXC depends on the assumption that on average the Malmquist bias for MCXC samples is the same as that of REXCEXX sample. 

\subsection{Cluster mass from the Sunyaev–Zel'dovich Effect}
In the Planck catalog of galaxy clusters \citep{Planck}, the mass ($M500$) for 439 galaxy clusters are estimated using gravitational lensing.

In Fig. 5 we show a scatter plot of distance and mass, from the Planck, MCXC, GWGC, RASS-SDSS, RASS-Abell catalogs.

\begin{figure}[htp]
\centering
\includegraphics[width=9cm]{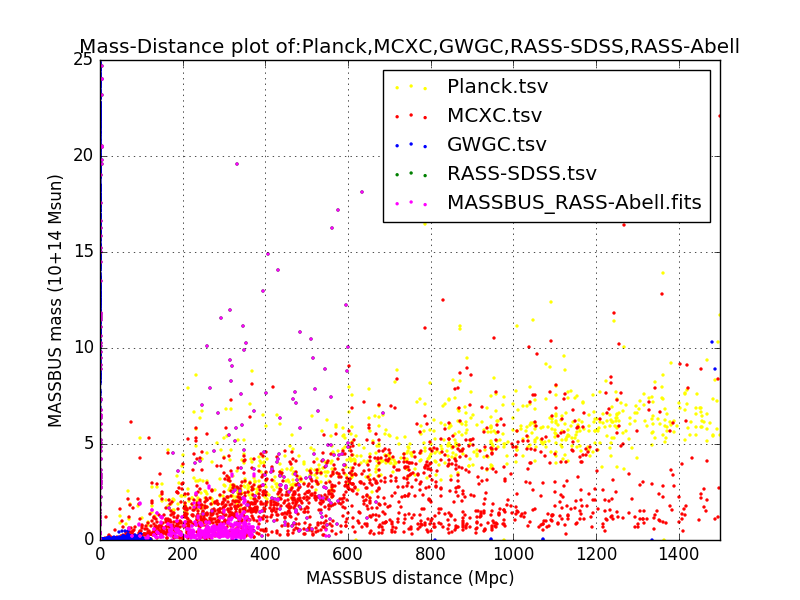}
\caption{Distance-mass plot of Planck, MCXC, GWGC ,RASS-SDSS, and RASS-Abell is shown in the plot for comparison. The stellar masses in GWGC are estimated by using blue luminosities and $M_\sun/L_\sun$. }
\end{figure}

\section{Metallicity Estimation}

Strategic searches for the EM counterpart have been anticipated for BNS and BHNS mergers, but for BBH mergers alone in vacuum, no obvious EM counterpart is expected. However, there are several scenarios, in which BBH mergers are expected to generate EM counterparts.
For example, long-lived disks with small masses may be formed around BHs \citep{Perna}. Also, a BBH in a hierarchical three-body system may trap a star and its tidal disruption can generate EM signals around the BBH merger \citep{seto}. The EM counterpart can be possibly generated by super-Eddington accretion onto BBHs in active galaxies, and also by the afterglow from relativistic jets.
Murase et al.\citep{murase} estimated the magnitude of possible EM counterparts of BBH mergers with mini-accretion disks around both BHs in the system. In their 2016 paper, they considered several different kinds of EM counterparts, including fast optical transients caused by disk-driven outflows, long-lasting radio emission from blast waves generated by ultra fast outflows, and possible after glow due to jet emission \citep{murase}. According to their calculation, the peak bolometric luminosity of a optical transient can reach $\sim 22+5log(d/100 \mathrm{Mpc})$ magnitude, and can last for hours to days, while the radio emission can last for years. Thus, it may be possible to observe EM counterparts of BBH mergers.

When building a prior for high-mass BBH, such as the $\sim 62 M_\sun$ BBH coalescence that was LIGO's first detection, metallicity becomes important. This is because stellar formation models cannot produce such high-mass black holes except from stars that formed from almost pure hydrogen (low metallicity) \citep{Belczynski}. Thus a galaxy of low metallicity gets a higher astrophysical prior than a high-metallicity galaxy of the same mass. 
Also, in low metallicity environment, rotational mixing tends to happen, and enable the third star in a hierarchal three-body system to survive the post-MS expansion of the BBH progenitor stars in the system. In this case, the accretion disk around the BBH due to the survived third companion star can catalyze the merger and give rise to a EM counterpart (Qingwen Lin). Due to the importance of metallicity in the prior of massive BBHs, we provide metallicity for each galaxy along with the stellar mass in Skymap Viewer.

For all GLADE-2MASS galaxies, metallicities are estimated together with stellar mass using the BayeSED. Metallicities of WISExSCOS galaxies, on the other hand, are derived from stellar mass using the empirical mass-metallicity relation \citep{ma}:
\begin{equation*}
    \log(Z_{gas}/Z_\sun)=0.35(\log(M_{gal})-10)+0.93e^{-0.43z}-1.05
\end{equation*}
In this calculation, we assume that the metallicity of a galaxy is uniform and equals to the mean metallicity of the star forming gas in the galaxy. The mass-metallicity relation comes from high-resolution cosmological simulation suite FIRE \citep{fire} , and it agrees with both gas and stellar metallicity measurements observed at low redshifts for $ 10^4 \leqslant M_{gal} \leqslant 10^{11} M_\sun $ \citep{tremonti}\citep{Lee} as well as the data at higher redshifts \citep{erb}\citep{mannuci}\citep{lamberts} .

\section{Summary}

The objective of this project is to expose a number of published catalogs of astrophysical sources into an interactive guide for follow-up observers of GW events. We have built plausible astrophysical priors for each source as a progenitor of GW signals, and rendered that information into Skymap Viewer. This prior, in principle, depends on the type of GW event being pursued, whether it is binary neutron star (where big, young stars are the progenitors), or binary black hole (where simple mass is the best prior). This latter case splits into lower and higher mass systems, because it is though that higher mass systems can only be produced in low-metallicity environments. However, given the lack of solid knowledge of the formation scenarios for the GW coalescences, we have not attempted to build different priors for different theories, but rather concentrated on two attributes that everyone can agree on: Mass and Distance. The distance is used with the new information from LIGO, that now provides 3D localization for the detected events.

To calibrate these mass estimates for galaxies, we have compared them by cross-matching the all-sky surveys that we want to use, with a much smaller, but more accurate survey (SDSS Stripe 82). Further, we have extended the GLADE-2MASS galaxy catalog by using SED (Spectral Energy Distribution) fitting to compute mass from colors.

The Appendix of this paper lists the catalogs used in Skymap Viewer, which attributes from the database table are used -- in addition to RA/Dec of course -- and the function used to get mass and distance from those attributes.

\bibliographystyle{apj}

\pagebreak
\appendix
For each of the catalogs shown in Skymap Viewer, we show in the table the Vizier code, the names of attributes that are extracted from the table, and the way that these are used to compute mass and distance. In the case of WISE$\times$SCOS, the catalog data comes from \citep{wisexscos}. In the case of GLADE-2MASS, the mass is computed as described in section 2.2.
\\
$H = 70$ km/s/Mpc \\
$c = 300000$ km/s \\
$h(z) = (0.3*(1+z)^3 + 0.7)^{1/2}$ \\

\begin{table*}[h]
\centering
\def\arraystretch{1.5}
\begin{tabular}{ l | l | l | l | l | r |}
Catalog & Vizier code & Attributes & Mass/$(10^{14} M_{\sun})$ & Distance/Mpc  \\
  \hline
GLADE-2MASS & VII/275      & Bmag,Dist& SED-fitting                          & $Dist$\\

GWGC & VII/267             & Bmag,Dist& $0.00855 Dist^2 10^{-0.4*Bmag}$      & $Dist$\\

WISExSCOS & (not in Vizier)& W1,W2,z  & $0.0018 d^2 10^{-2.36 W1 + 1.96 W2}$ & $d = cz/(Hh(z))$ \\

MCXC & J/A+A/534/A109      & M500,z   & $M500$                               & $c z/(Hh(z))$ \\

Planck & J/A+A/581/A14     & MYZ500,z & $MYZ500$                             & $c z/(Hh(z))$ \\

RASS-Abell & J/AJ/126/2740 & LX,z     & $0.501 [h(z)^{-7/3} LX]^{0.610}$     & $c z/(Hh(z))$ \\

RASS-SDSS & J/A+A/461/397  & M200,z  & $M200$                                & $c z/(Hh(z))$ \\

\end{tabular}
\end{table*}

\end{document}